\documentclass{emulateapj}
\usepackage{apjfonts}

\usepackage{amsmath}

\usepackage{graphics}

\newcommand{\bd}{\begin{displaymath}}
\newcommand{\ed}{\end{displaymath}}

\slugcomment{accepted by ApJ Letters}

\shorttitle{How many radio quasars can be detected by GLAST?}

\begin{document}

\title{How many radio-loud quasars can be detected by the {\it
Gamma-Ray Large Area Space Telescope}?}

\author{Xinwu Cao\altaffilmark{1} and J. M. Bai\altaffilmark{2}}

\begin{abstract}

In the unification scheme, radio quasars and FR II radio galaxies
come from the same parent population, but viewed at different
angles.  Based on the Comptonization models for the $\gamma$-ray
emission from active galactic nuclei (AGNs), we estimate the number
of radio quasars and FR II radio galaxies to be detected by the {\it
Gamma-Ray Large Area Space Telescope} (GLAST) using the luminosity
function (LF) of their parent population derived from the
flat-spectrum radio quasar (FSRQ) LF. We find that $\sim1200$ radio
quasars will be detected by GLAST, if the soft seed photons for
Comptonization come from the regions outside the jets. We also
consider the synchrotron self-Comptonization (SSC) model, and find
it unlikely to be responsible for $\gamma$-ray emission from radio
quasars. We find that no FR II radio galaxies will be detected by
GLAST. Our results show that most radio AGNs to be detected by GLAST
will be FSRQs ($\sim$99\% for the external Comptonization model, EC
model), while the remainder ($\sim$1\%) will be steep-spectrum radio
quasars (SSRQs). This implies that FSRQs will still be good
candidates for identifying $\gamma$-ray AGNs even for the GLAST
sources. The contribution of all radio quasars and FR II radio
galaxies to the extragalactic $\gamma$-ray background (EGRB) is
calculated, which accounts for $\sim$30\% of the EGRB.

\end{abstract}


\altaffiltext{1}{Shanghai Astronomical Observatory, Chinese Academy
of Sciences, 80 Nandan Road, Shanghai, 200030, China;
cxw@shao.ac.cn} \altaffiltext{2}{National Astronomical
Observatories/Yunnan Observatory, Chinese Academy of Sciences, P.O.
Box 110, Kunming, Yunnan 650011, China}

\section{Introduction}

The third catalog of $\gamma$-ray AGNs detected by the Energetic
Gamma-Ray Experiment Telescope (EGRET) on the {\it Compton Gamma-Ray
Observatory (CGRO)} includes $\sim$ 80 high-confidence
identifications of blazars \citep[e.g.,][]{h99,ma01}. GLAST has
higher sensitivity than EGRET, and much more blazars are expected to
be detected after its launch. Many workers have predicted the
statistic properties of blazars in the GLAST era
\citep*[e.g.,][]{s99,p07,d07}. One method is to extrapolate the
observed $\gamma$-ray luminosity distribution of EGRET blazars to
obtain a $\gamma$-ray luminosity function (LF) \citep{c95}. An
alternative method is to assume some correlation between
$\gamma$-ray emission and the emission in other bands to model the
undetected $\gamma$-ray blazars, in which the larger samples in
other bands provide useful clues to such researches
\citep*[e.g.,][]{p93,s93,d07,p07}. The previous works on the EGRB
showed that about $\sim 25\%$ to $\sim 100\%$ of the EGRB can be
attributed to the unresolved blazars
\citep*[e.g.,][]{p93,c95,ss96,mp00,nt06}.

Comptonization is widely believed to  be responsible for the
$\gamma$-ray emission from the blazars detected by EGRET, which can
be classified into two categories: the EC models and SSC model,
according to the origin of the soft seed photons \citep[see
e.g.,][for a review and references therein]{b07}. The space density
and evolution of the parent population of blazars, together with the
Lorentz factor distribution of the jets, are crucial for
understanding the properties of $\gamma$-ray emitting blazars. In
almost all previous works, the models of blazars are rather
simplified. In this {\it Letter}, we derive the parent radio LF of
radio quasars/FR IIs from the FSRQ LF to investigate the statistic
properties of $\gamma$-ray emitting quasars to be detected by GLAST.
The cosmological parameters $\Omega_{\rm M}=0.3$,
$\Omega_{\Lambda}=0.7$, and $H_0=70~ {\rm km~s^{-1}~Mpc^{-1}}$ have
been adopted in this {\it Letter}.

\section{EC and SSC models for $\gamma$-ray AGNs}

In the EC models, the observed $\gamma$-ray emission from the
relativistic jet is closely related to its observed radio emission
\citep*[see Eq. 26 in][]{dss97},
\begin{equation}
\nu L_{\nu,\gamma}^{\rm EC}\simeq{\frac {3u^*_{\rm i}}{4(p+3)u_{\rm
H}}} {\frac {(1+\mu_{\rm obs})^{(p+3)/2}}{\mu_{\rm obs}}}
\delta_{\rm j}^{(p+1)/2}\nu L_{\nu,{\rm j}}^{\rm rad},
\label{l_ecrad}
\end{equation}
where $u_{\rm i}^*$ is the soft seed photon energy density measured
in the stationary source frame, $u_{\rm H}$ is the magnetic energy
density in the jet, $\mu_{\rm obs}=\cos\theta_{\rm obs}$
($\theta_{\rm obs}$ is the direction of the jet motion with respect
to the line of sight), $\delta_{\rm j}=[\gamma_{\rm j}(1-\beta_{\rm
j}\mu_{\rm obs})]^{-1}$ for a jet moving at $\beta_{\rm j}c$, and
$\nu L_{\nu,{\rm j}}^{\rm rad}$ is the observed radio luminosity of
the jet. The energy distribution of the nonthermal electrons in the
jet is assumed to be $n_{\rm e}\propto \gamma_{\rm e}^{-p}$.

In the EC models, the soft photons may originate from the accretion
disks, the broad-line regions (BLRs), or/and the dust tori
\citep*[e.g.,][]{gm96,gkm01,ds02}. It was argued that the
contribution from the accretion disks is not important, because the
$\gamma$-ray emitting region is far away from the disk and in the
jet comoving frame, the energy density of photons from the disk is
deboosted by the relativistic jet moving away from the black hole
\citep*[e.g.,][]{sbr94,dss97}. It is well known that the BLR size
$R_{\rm BLR}\propto L_{\rm bol}^{\alpha_{\rm BLR}}$, where
$\alpha_{\rm BLR}\simeq 0.5-0.7$ \citep*[e.g.,][]{k00,b06}.
\citet{b06} found that $\alpha_{\rm BLR}$ is 0.518 subtracting the
contribution from the host galaxy starlight to $L_{\rm bol}$, which
is consistent with $\alpha_{\rm BLR}\simeq 0.5$ expected from the
photo-ionization model if all BLRs have similar physical properties.
The inner radius of the dust torus is roughly at the dust
evaporation radius: $R_{\rm inn}\propto L_{\rm bol}^{0.5}$
\citep{nl93}. The photon energy density $u_{\rm i}^*\propto L/R^2$,
where $L=L_{\rm BLR}$ and $R=R_{\rm BLR}$ for the BLR photons, and
$L=L_{\rm IR}$ and $R\sim R_{\rm inn}$ for the dust torus case. The
irradiated infrared luminosity of the dust torus $L_{\rm IR}\propto
L_{\rm bol}$, if the opening angle of the torus does not vary much
for individual sources \citep*[e.g.,][]{cao05}. Thus, the energy
density of the soft photons from the BLRs/dust tori is roughly
universal for most sources. We rewrite Eq. (\ref{l_ecrad}) as
\begin{equation}
\nu L_{\nu,\gamma}^{\rm EC}={\cal C_{\rm EC}} {\frac {(1+\mu_{\rm
obs})^{(p+3)/2}}{\mu_{\rm obs}}} \delta_{\rm j}^{(p+1)/2}\nu
L_{\nu,{\rm j}}^{\rm rad}, \label{l_ecrad2}
\end{equation}
where the normalization $\cal C_{\rm EC}$ is related with $u^*_{\rm
i}/u_{\rm H}$ (see Eq. \ref{l_ecrad}). For the SSC model, the
observed $\gamma$-ray luminosity is \citep*[see Eq. 28 in][]{dss97}
\begin{equation}
\nu L_{\nu,\gamma}^{\rm SSC}={\cal C_{\rm SSC}}\nu L_{\nu,{\rm
j}}^{\rm rad}, \label{l_sscrad}
\end{equation}
where $\cal C_{\rm SSC}$ is the normalization.

\section{The parent LF of FSRQs}

In the unification scheme, FRSQs, SSRQs, and FR II galaxies come
from the same parent population, but viewed at different angles.
Some previous authors have derived the parent LF using different
approaches \citep*[see][]{cl07,lz07}, however, they have not
compared the number density of blazars with that of radio galaxies.

\citet{pu92} derived the radio LFs of FSRQs and FR II galaxies from
a sample of radio-loud AGNs. They considered a two-component model,
in which the total luminosity $L_{\nu,\rm T}$ is the sum of an
unbeamed part $\cal L_{\nu,\rm u}$ and a jet luminosity
\begin{equation}
L_{\nu,\rm j}=\delta_{\rm j}^{3+\alpha_{\rm rad}}\cal L_{\nu,\rm j}.
\label{ljdelta}
\end{equation}
They used a variety of observational features to constrain the ratio
$f$($\equiv \cal L_{\nu,\rm j}/\cal L_{\nu,\rm u}$). They found that
a constant $f\simeq 4.5\times 10^{-3}$ can successfully explain the
observations \citep*[see][for the details]{pu92}. Thus, FSRQs should
satisfy $\delta_{\rm j}> \delta_{\rm j,min}\simeq 6.45$, as their
core dominance parameters  $R=L_{\nu,\rm j}/{\cal L}_{\nu,\rm u}\ga
1$ is required, where an average core spectral index $\alpha_{\rm
rad}=-0.1$ is adopted.  They further assumed that the probability
distribution of the Lorentz factors for the jets is $P(\gamma_{\rm
j})=C\gamma_{\rm j}^G$, between $\gamma_{\rm j,1}=5$ and
$\gamma_{\rm j,2}=40$. Their derived FSRQ LF is consistent with the
beaming model and the LF of FR II galaxies, provided $G=-2.3$ is
adopted. Recently, \citet{pg07} derived a FSRQ LF based on the deep
X-ray radio blazar survey (DXRBS) in the same way, which extends to
lower luminosity than that derived by \citet[][]{pu92}.

The sources in the parent population may be observed as FR IIs, when
their jets are oriented at angles $\theta_{\rm obs}\ga40^\circ$ to
the line of sight \citep*[e.g.,][]{pu92}. The sources in this parent
population with $\theta_{\rm obs}\la 40^\circ$ and $\delta_{\rm
j}<\delta_{\rm j,min}$ will appear as SSRQs. The LFs of FSRQs,
SSRQs, or FR II galaxies can be reproduced with this parent radio
LF, if the probability distribution of the Lorentz factors
$P(\gamma_{\rm j})$ is supplied. The LF of FSRQs $\phi_{\rm
FSRQ}(L_{\nu,\rm j})$ can be derived from the LF of the parent
population $\phi(\cal L_{\nu,\rm j})$ by
\begin{equation}
\phi_{\rm FSRQ} (L_{\nu,\rm j},z) =\int\limits_{\gamma_{\rm
j,1}}^{\gamma_{\rm j,2}} P(\gamma_{\rm j}){\rm d}\gamma_{\rm j}\int
\limits_{\mu_{\rm obs}^{\rm min}(\gamma_{\rm j})}^{1} \phi({\cal
L_{\nu,\rm j}},z){\frac {{\rm d}{\cal L}_{\nu,\rm j}}{{\rm
d}L_{\nu,\rm j}}} {\rm d}\mu_{\rm obs}, \label{parentlf}
\end{equation}
where the orientations of the jets of the parent population are
assumed to be isotropically distributed, and only those with
$\mu_{\rm obs}\ge\mu_{\rm obs}^{\rm min} (\gamma_{\rm
j})=(\gamma_{\rm j}\delta_{\rm j,min}-1)/(\gamma_{\rm
j}^2-1)^{1/2}\delta_{\rm j,min}$ are FSRQs, which is required by
FSRQs having $R\ga 1$. Equation (\ref{parentlf}) can be re-written
as
\begin{equation}
\phi_{\rm FSRQ} (L_{\nu,\rm j}^i,z) =\sum_{k=1}^{n}\phi({\cal
L}_{\nu,\rm j}^k,z)\epsilon_{ik}, \label{parentlf_num}
\end{equation}
where $i=1,n$, and
\begin{equation}
\epsilon_{ik}= \int\limits_{\gamma_{\rm j,1}}^{\gamma_{\rm j,2}}
P(\gamma_{\rm j}){\rm d}\gamma_{\rm j}\int \limits_{\mu_{\rm
obs}(\gamma_{\rm j})}^{} \left.{\frac {{\rm d}{\cal L}_{\nu,\rm
j}}{{\rm d}L_{\nu,\rm j}}}\right |_{ {L}_{\nu,\rm j}={L}_{\nu,\rm
j}^i} {\rm d}\mu_{\rm obs}. \label{epsilonik}
\end{equation}
The coefficient $\epsilon_{ik}$ can be calculated with Eq.
(\ref{epsilonik}) by using Eq. (\ref{ljdelta}) and adopting the
integral limits $\mu_{\rm obs}(\gamma_{\rm j})$ to satisfy ${\cal
L}_{\nu,\rm j}^k-\Delta {\cal L}_{\nu,\rm j}/2\le {\cal L}_{\nu,\rm
j} \le {\cal L}_{\nu,\rm j}^k+\Delta {\cal L}_{\nu,\rm j}/2$.
Solving a set of $n$ linear algebraic equations (\ref{parentlf_num})
numerically, the parent LF $\phi({\cal L}_{\nu,\rm j},z)$ can be
calculated from the LF of FSRQs $\phi_{\rm FSRQ} (L_{\nu,\rm j},z)$
given by \citet{pg07}.

\section{Number of GLAST quasars}

Using this derived parent radio LF, we can calculate the observed
$\gamma$-ray LF $\phi_\gamma$ for FSRQs, SSRQs and FR II galaxies
based on either EC or SSC models:
\begin{equation}
\phi_\gamma^{\rm EC/SSC}(L_{\nu,\gamma}^{\rm EC/SSC},z)
=\int\limits_{\gamma_{\rm j,1}}^{\gamma_{\rm j,2}} P(\gamma_{\rm
j}){\rm d}\gamma_{\rm j} \int\limits_{\mu_{\rm obs}(\gamma_{\rm
j})}^{} \phi({\cal L_{\nu,\rm j}},z){\frac {{\rm d}{\cal L}_{\nu,\rm
j} } {{\rm d}L_{\nu,\gamma}^{\rm EC/SSC}} } {\rm d}\mu_{\rm obs},
\label{gammalf}
\end{equation}
where ${{\rm d}\cal L_{\nu,{\rm j}}}/{{\rm d}L_{\nu,\gamma}^{\rm
EC/SSC}}$ can be derived from Eqs.
(\ref{l_ecrad2})$-$(\ref{ljdelta}). This $\gamma$-ray LF is not
limited to blazars, as their parent radio LF is adopted in the
calculations. Adopting the conditions for different sources (i.e.,
$\delta_{\rm j}> \delta_{\rm j,min}$ for FSRQs; $\theta_{\rm obs}\ga
40^\circ$ for FR II galaxies; and the remainder are SSRQs), the
numbers of the FSRQs/SSRQs/FR II galaxies with $f_{\nu,\gamma}\ge
f_{\nu,\gamma}^{\rm min}$ as functions of redshift $z$ can be
calculated by
\begin{equation}
{\frac {{\rm d}N^{\rm EC/SSC}(z)}{{\rm d}z}} =\int\limits_{4\pi
d_{\rm L}^2f_{\nu,\gamma}^{\rm min}}^{} {\frac {{\rm d}V}{{\rm d}z}}
\phi_\gamma^{\rm EC/SSC}(L_{\nu,\gamma}^{\rm EC/SSC},z){\rm
d}L_{\nu,\gamma}^{\rm EC/SSC}, \label{dndz}
\end{equation}
either for EC or SSC models, respectively.

Sixty three $\gamma$-ray emitters were identified as blazars at high
confidence with measured redshifts in \citet{h99}. More $\gamma$-ray
sources were identified as blazars afterwards
\citep{ma01,s-e03,s-e04,sg04}. We collect all these $\gamma$-ray
blazars identified at high confidence, which leads to 64 quasars and
16 BL Lac objects with measured redshifts. We note that the measured
fluxes with photon energy greater than 100~MeV $\ga 5\times 10^{-8}$
photons~cm$^{-2}$ for all EGRET blazars. This can be translated to
${\nu}f_{\nu,\rm EGRET}^{\rm min}\simeq 8\times 10^{-12}$
erg~s$^{-1}$~cm$^{-2}$ at 100~MeV assuming a mean photon spectral
index of 2 \citep*[][]{h99}, which corresponds to $p=3$
\citep{dss97}.

We calculate the number counts of $\gamma$-ray quasars with
$\gamma$-ray flux densities greater than $f_\nu^{\rm min}$ at
100~MeV using Eq. (\ref{dndz}), based on the different $\gamma$-ray
radiative models (EC or SSC models). We tune the values of the
parameters ${\cal C}_{\rm EC}$ or ${\cal C}_{\rm SSC}$ to let the
total number of FSRQs derived with Eq. (\ref{dndz}) equal to that of
the FSRQs detected by EGRET. We find that ${\cal C}_{\rm
EC}=4.42\times 10^{-3}$ or ${\cal C}_{\rm SSC}=10.62$ are required
to reproduce 64 FSRQs detected by EGRET for the EC or SSC models,
respectively. Based on the derived values of ${\cal C}_{\rm EC}$ or
${\cal C}_{\rm SSC}$, the number counts of $\gamma$-ray emitting
quasars/FR IIs to be detected by GLAST can be predicted by adopting
different flux density limits $f_{\nu,{\rm min}}$ with Eq.
(\ref{dndz}) (see Fig. \ref{fig1} and Table 1).

\vskip 1.0cm \figurenum{1}
\centerline{\includegraphics[angle=0,width=9.0cm]{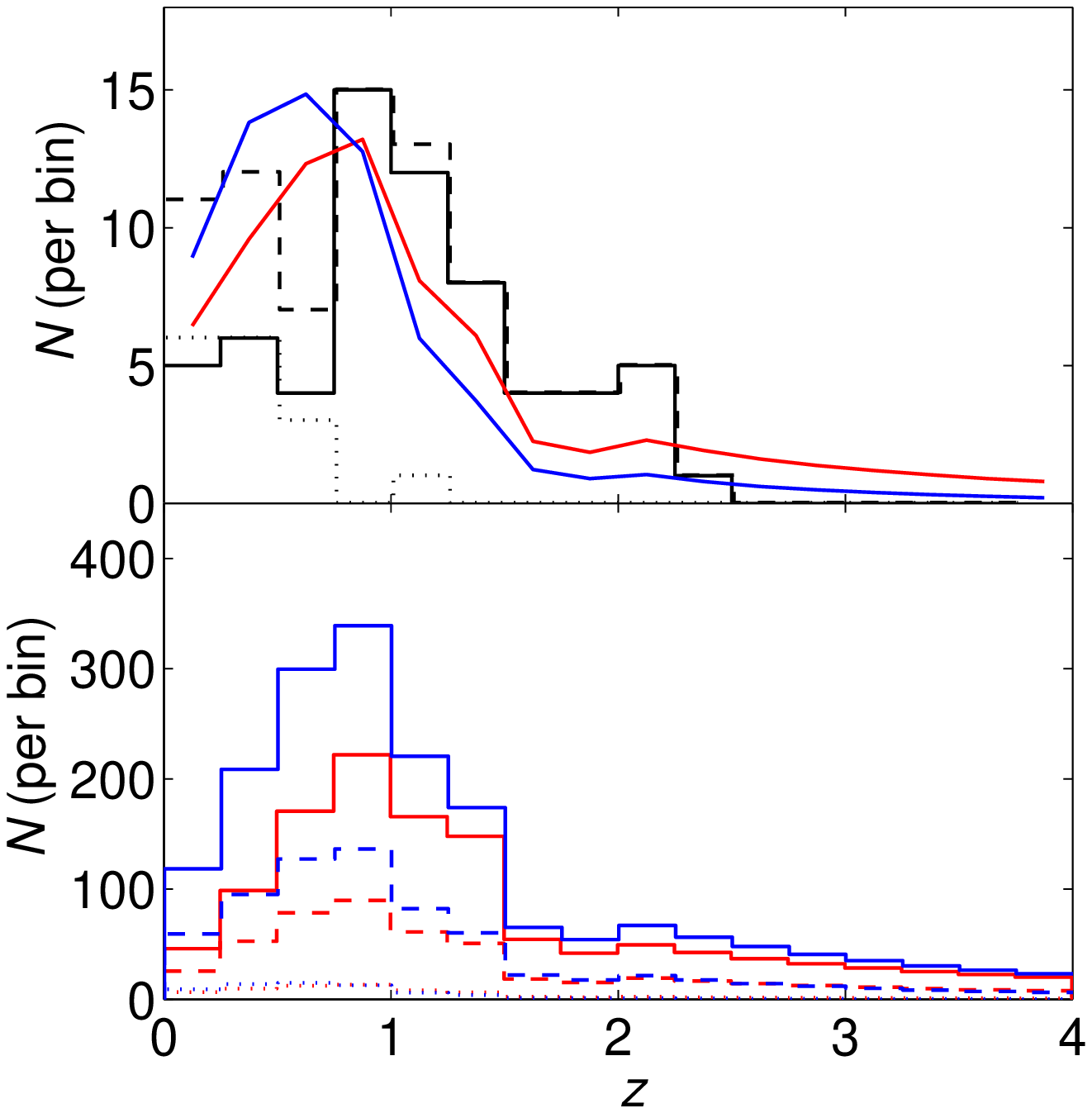}}
\figcaption{Upper panel: the redshift distributions of the EGRET
blazars (the black solid line: quasars; the dashed line: all EGRET
blazars; the dotted line: BL Lac objects). The color lines represent
our model calculations for 64 quasars. The red line represents the
model prediction based on the EC model, while the blue line
represents the result for SSC model. Lower panel: the calculated
redshift distributions of $\gamma$-ray quasars detected with
different flux density limits. The red lines represent the
calculations based on the EC model, while the blue lines are for the
SSC model. The dotted lines represent our model calculations for
EGRET quasars. The dashed lines represent the redshift distribution
of the $\gamma$-ray quasars with $\ge 0.1f_{\rm limit}^{\rm EGRET}$
at 100~MeV, while the solid lines are for the sources with $\ge 1/30
f_{\rm limit}^{\rm EGRET}$. \label{fig1} }\centerline{}

The total numbers of $\gamma$-ray emitting radio quasars/FR IIs as
functions of sensitivity are plotted in Fig. \ref{fig2}. About 1200
$\gamma$-ray radio quasars will be detected by GLAST based on the EC
model, if its sensitivity is 30 times higher than that of EGRET at
100~MeV \citep[][]{gm99}. Our calculations show that no FR II radio
galaxies will be detected by GLAST as $\gamma$-ray emitters either
for EC or SSC models. We find that almost all $\gamma$-ray quasars
($\sim$99\%) to be detected by GLAST will be FSRQs for the EC model,
and the remainder ($\sim$1\%) will be SSRQs. For the SSC model,
$\sim 1800$ quasars will be detected by GLAST,  of which $\sim$80\%
will be FSRQs (see Fig. \ref{fig2}). We use the derived $\gamma$-ray
LF (Eq. \ref{gammalf}) to calculate the contribution of all radio
quasars/FR IIs to the EGRB (listed in Table 1).

\vskip 1.0cm \figurenum{2}
\centerline{\includegraphics[angle=0,width=9.0cm]{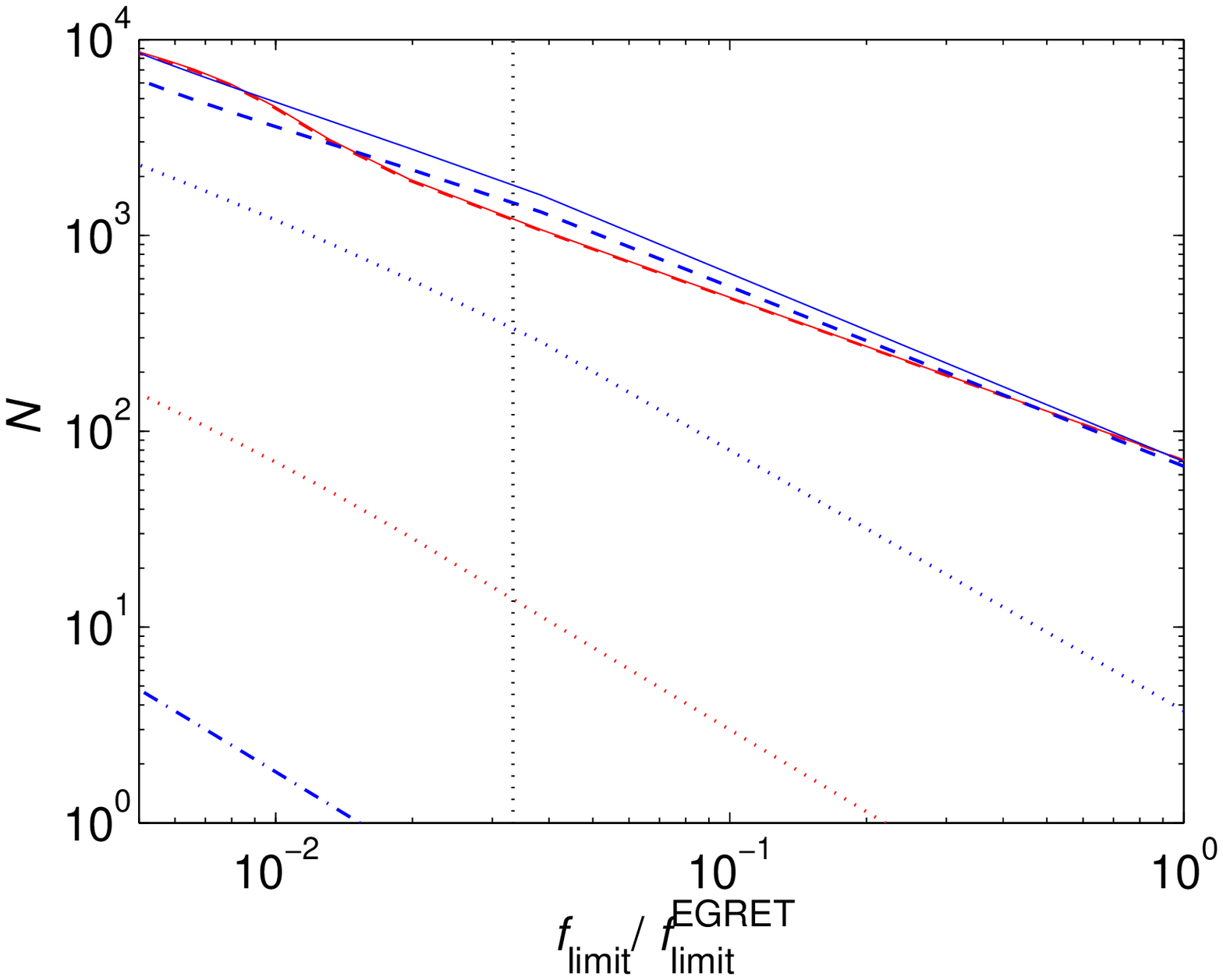}}
\figcaption{The total number of the $\gamma$-ray quasars with flux
densities greater than $f_{\rm limit}$ at 100~MeV. The red lines are
calculated for the EC model, while the blue lines are for the SSC
model. The dashed lines represent the $\gamma$-ray FSRQs, while the
colored dotted lines represent the $\gamma$-ray SSRQs. The
dash-dotted line represents the FR II radio galaxies. The black
dotted line represents the GLAST sensitivity at 100 MeV.
\label{fig2} }\centerline{}

\section{Discussion}

We find that the redshifts of almost all EGRET BL Lac objects are
$\la 1$, which implies that BL Lac objects may have different space
density and evolutionary behaviors from quasars. The LF of BL Lac
objects was derived from the DXRBS by \citet{pg07}, however, the
results for BL Lac objects are more uncertain than those for FSRQs,
because of the small number statistics and $\sim$30\% of them having
no redshift. In this work, we use a parent radio LF of radio
quasars/FR II galaxies derived from the FSRQ LF. The derived
redshift distributions of $\gamma$-ray quasars are similar for
different models (EC or SSC), which are roughly consistent with that
of the EGRET quasars.

It was suggested that the $\gamma$-ray radiative mechanisms are
different for quasars and BL Lac objects, i.e., the EC mechanism may
be responsible for quasars, while the SSC is for BL Lac objects
\citep*[e.g.,][]{dg95}. If the EC mechanism is indeed responsible
for $\gamma$-ray quasars, the predicted $\gamma$-ray quasars to be
detected by GLAST will be $\sim1200$. Our results are roughly
consistent with the estimate given by \citet{d07} based on a
simplified blazar model. The SSC model predicts a simple relation
between $\gamma$-ray luminosity and radio luminosity of the jets
(see Eq. \ref{l_sscrad}). The EGRET flux limit ${\nu}f_{\nu,\rm
EGRET}^{\rm min}$ at 100~MeV can be converted to a radio flux
density limit $f_{\nu,\rm rad}\sim 10$~Jy at 5GHz by using Eq.
(\ref{l_sscrad}). This means that all EGRET quasars should have
their radio flux densities higher than $\sim 10$ Jy, which is
obviously inconsistent with most EGRET quasars having $f_{\nu,\rm
rad}\ga 1$~Jy \citep*[e.g.,][]{s93,zhou97}. Thus, the SSC model is
unlikely to be responsible for EGRET quasars, unless the physical
properties of the jets are significantly different for individual
sources, i.e., the values of ${\cal C}_{\rm SSC}$ for most sources
deviate significantly from a constant value.

Our results show that no FR II galaxies will be detected by GLAST.
Most GLAST quasars will be FSRQs ($\sim$99\% for the EC model),
which implies that FSRQs will still be good candidates for
identifying the $\gamma$-ray sources even for the GLAST sources. At
least two FR I galaxies have been detected by EGRET
\citep*[e.g.,][]{p07}, and more FR I galaxies were predicted to be
detected by GLAST \citep*[e.g.,][]{gtc05}. In the unification
scheme, FR I galaxies are BL Lac objects with misaligned jets. The
$\gamma$-ray FR I galaxies/BL Lac objects are beyond the scope of
this work. The EGRET quasars identified at high confidence (64
sources) have radio flux densities $\ga 1$~Jy and are about 20\% of
the total FSRQs ($\sim 300$) above the same flux density limit
\citep*[see Fig. 6 in][]{pg07}. Assuming the radio flux density
limit of the GLAST quasars to be $\sim 30$ times lower than the
EGRET limit, i.e., $\sim 30$~mJy, the total all-sky number of the
FSRQs above this limit is about $20000$. So, about 4000 FSRQs will
be detected by GLAST, if the same percentage ($\sim 20$\%) is
adopted as the EGRET blazars. This rough estimate is about a factor
of two higher than our model calculations. Considering that some
unidentified $\gamma$-ray sources are likely to be blazars, our
estimates on $\gamma$-ray quasars to be detected by GLAST are only
lower limits. Both the duty cycle in the $\gamma$-ray band and
identification rate may affect the detection rate, and we implicitly
assume them to be similar to those of EGRET blazars.

The EGRB integrated above 100 MeV was determined to be $1.45(\pm
0.05)\times 10^{-5}$ photons cm$^{-2}$~s$^{-1}$~sr$^{-1}$ from the
EGRET data \citep{s98}. \citet{s04} used a new model of the Galactic
background, and obtained a slightly smaller value of the EGRB,
$1.14(\pm 0.12)\times 10^{-5}$ photons cm$^{-2}$~s$^{-1}$~sr$^{-1}$.
We sum up the $\gamma$-ray emission from all radio quasars/FR IIs
with our derived $\gamma$-ray LF and find that they contribute
$\sim$30\% of the EGRB (see Table 1), which are only lower limits,
because our calculations are limited to radio quasars/FR II
galaxies. The BL Lac objects/FR Is must contribute some part to the
EGRB. The detailed calculation of their contribution to the EGRB is
beyond the scope of this {\it Letter}.

\acknowledgments  We thank the anonymous referee for his/her helpful
comments/suggestions, and Paolo Padovani for providing us the data
of the LF of FSRQs. This work is supported by the NSFC (grants
10325314, 10333020, 10573030 and 10773020), and the CAS (grant
KJCX2-YW-T03).

{}


\begin{deluxetable}{cccccc}
\tabletypesize{\scriptsize} \tablecaption{The summary of different
models} \tablewidth{0pt} \tablehead{ \colhead{${\cal C}_{\rm EC}$} &
\colhead{${\cal C}_{\rm SSC}$} & \colhead{N($\ge {{\nu}f_{\nu,\rm
limit}^{\rm EGRET}}$)$^{\rm a}$} & \colhead{N($\ge
1/10{{\nu}f_{\nu,\rm limit}^{\rm EGRET}}$)} & \colhead{N($\ge
1/30{{\nu}f_{\nu,\rm limit}^{\rm EGRET}}$)} & \colhead{$f_{\rm
EGRB}^{\rm b}$} } \startdata
4.42$\times 10^{-3}$ & ... & 64 & 491 & 1203 &  0.32 \\
...  & 10.62 & 64 & 696 & 1806 & 0.30 \\
\enddata
 \tablenotetext{a}{The number of FSRQs with $\ge {\nu}f_{\nu,\rm limit}^{\rm EGRET}\simeq 8\times 10^{-12}$ erg~s$^{-1}$~cm$^{-2}$ at
100~MeV.}; \tablenotetext{b}{for the photons with energy above
100~MeV in units of 10$^{-5}$ photons~cm$^{-2}$~s$^{-1}$~sr$^{-1}$.}

\end{deluxetable}

\end{document}